\documentclass[%
 reprint,
 amsmath,amssymb,
 aps,
]{revtex4-1}

\usepackage[utf8]{inputenc}
\usepackage{amsthm}

\theoremstyle{definition}
\newtheorem{definition}{Definition}[section]
\usepackage{csquotes}
\usepackage{graphicx}
\usepackage{dcolumn}
\usepackage{bm}

\begin{document}


\title{A Theory of Decision Making Based on Feynman's Path Integral Formulation of Quantum Mechanics}

\author{Peter G. Solazzo}
\affiliation{Johns Hopkins University Applied Physics Laboratory}

\date{\today}

\begin{abstract}
Similarities between the non-deterministic nature of quantum theory and the unpredictable patterns of human cognition and decision making have been observed and commented on many times since the invention of Quantum Mechanics in the first part of the $20^{th}$ century.  Niels Bohr himself took note of the parallels.\cite{ashtianai}  In fact, an entire field of study, Quantum Cognition, has been borne from the study of this analogy. \cite{ashtianai}  However, many of the attempts to model human behavior with quantum mechanics conflate the identity of a particle with its own wavefunction, which is incorrect and invalidates the analogy.  In this paper, we seek to make explicit this error, make necessary corrections, and then deepen the analogy. We do this by creating a Quantum Decision Theory that directly parallels Richard Feynman's novel formulation of quantum mechanics published in 1948 at Cornell University: ``Space-Time Approach to Non-Relativistic Quantum Mechanics''.
\end{abstract}

\maketitle


\section{\label{sec:level1}Introduction}

The field of quantum cognition hypothesizes that human behavior and decision making can best be modeled using the framework of quantum mechanics designed for physics in the late 20$^{th}$ century.  
In the framework of quantum mechanics, particles lose their deterministic properties, and instead obey laws of probability.
This loss of determinism is what makes the theory so attractive to those interested in modeling human behavior.
Humans tend not to be deterministic, or in many cases, even decisive.
The mathematical machinery of quantum mechanics is uniquely suited to approach just such a problem.

\par
However, there exists a popular misinterpretation of quantum theory wherein the ``state'' of a system, along with its associated wavefunction, is understood to refer to its measurable or physical manifestation.  
In this interpretation, the current ``state'', or wavefunction of the system, is a superposition of the many available future states that the system in question might be observed to take on.
In the context of decision making, the system is a mind.
Prior to making the decision, the state or wavefunction consists of a superposition of all of the possible decisions.
When the mind comes to a conclusion, the wavefunction is said to collapse into the decision actually made. 
 In this sense a decision amounts to a measurement.
 \par
 While intuitively attractive, this analogy to quantum mechanics contains a fundamental flaw.
 It assumes either that a quantum particle (in this case, a mind) is equivalent to its wavefunction or that its wavefunction can be measured.  
 More explicitly, we said that decisions are components of the wavefunction, and therefore are wavefunctions themselves.
 Yet, in the final analysis, we can ``measure'' the decision actually made, which would amount to measuring the wavefunction itself, and wavefunctions are not obeservables.
 Decisions cannot simultaneously be observables and operators without violating the machinery of quantum mechanics.
 A careful analysis of proper language reassures us of this conclusion.
 We say that a particle ``exists in a superposition of states''.  
 We do not say that a particle ``is a superposition of states''.  
 The latter statement would assert that a physical particle (or mind) is itself a linear combination of complex waveforms, which is not the case.
 \par
 Rather, particles are objects that have properties.  
 Particles have position, momentum, charge, spin, and many other interesting properties.
 The wavefunction is simply a mathematical tool for predicting the behavior of the particle given these properties.
 \par
 In what follows, we shall remake this analysis to reflect the statement that a particle ``exists in a superposition of states''.
 The fundamental analogy we shall draw is that \textit{the mind is a quantum mechanical particle, and therefore can be modeled using quantum mechanics}.
 This analogy is importantly different from the statement that the mind is a superposition of quantum states.
 Finally, mirroring Feynman's path-integral formulation of quantum mechanics, we shall formulate a quantum theory of decision making.
 
 \section{\label{sec:level1} Illustration of Conflating the Mind and Measureable Quantities with Wavefunctions and States}
 The error of attempting to analogize a mind to a waveform rather than a particle is anything but semantic.
 In order to achieve the desired behaviour of the model, the mind must be a particle not a waveform.
 Furthermore, if the mind is a waveform, it has no observables, and the theory collapses entirely.
 
 \subsection{\label{sec:level2} A First Example}
 As an example of this misconception at work, we examine a paper considering Quantum Decision Theory (QDT).
 This paper is not unique, it was simply the example chosen.
 \underline{Quantum field inspired model of decision making}, by Bagarello, Basieva, and Kherennikov, rightly and sharply distinguishes between a theory that requires actual quantum physical phenomena in the brain, and one which only uses the machinery of quantum theory to model human behaiour.
 Their theory and the theory presented in this paper are emphatically the latter.
\par
 However, from this point on, troubling statements follow.  
 ``In the simplest model, the mental state (the belief state) of an agent Alice, is represented as a quantum state $\psi$ and questions or tasks as quantum observables... Alice's decision is represented as a measurement of the observable $A$ at some instant in time'' \cite{bagarello}
 
 The first part of the statement, illustrates precisely the mistake highlighted earlier.
 Neither Alice, nor her ``belief state'' can accurately be called a quantum state.  
 Alice is an observable person, she exists in the real world.
 Her beliefs are observables. 
 One can ask Alice what she believes, presumably by operating with $\hat{A}$ to get the obervable $A$ as suggested above, and recieving a response corresponding to a particular belief.
 In this scenario, the wavefunction, which is, by definition, immeasurable has been measured.
 If what is stated in the paper were literally true, Alice's mind would simply not exist.
 \textit{The states are not real.  They are mathematical tools only.}
 This is an example of conflating the mind, an observable thing, with its wavefunction, which is not observable.
 
\par
Other deep misconceptions are related to this fundamental misunderstanding.
In the paper, Alice's environment also becomes a set of states.
Again, this is impossible as the environment is observable.
The root of the issue may be the use of the word state in quantum mechanics.

\section{\label{sec::level1} The Corrected Analogy Between Decision Making and Quantum Theory}
The fundamental analogy of our model is that a decision maker is like a quantum particle.
As particles exist in real spacetime, our decision maker will exist in decision space.
\theoremstyle{definition}
\begin{definition}{Decision Space: }
	The decision space is defined to be the set of all decisions available to the decision maker at any given time $t$.
\end{definition}

If the decision maker exists in decision space as a particle exists in spacetime, the next task is to determine the equations of motion of the decision maker in this space.
Here, as in Feynman's formulation of quantum mechanics, this is done by finding a Lagrangian.
In physics, Lagrangians expose the topology of the kinetic and potential energy around a particle.
Here we shall construct a similar topology for decision space.
Note carefully that we are not directly constructing the states.
Instead, the states will be calculated from the energetic topology of the decision space, defined by the Lagrangian.
Just as in real quantum mechanics, where one derives the states from the shape of the potential, not the other way around.

\section{\label{sec::level1} Using this Analogy to develop a Quantum Theory of Decision Making}
The following analysis will be agnostic to the exact form of the Lagrangan, so long as all terms are of order less than 2, for the reasons discussed by Feynman. \cite{feynman}
Extensions of this theory with perturbation analysis will also be valid.
For clarity, however, we shall do a quick exercise in constructing a Lagrangian for this type of problem, taking the Principle of Least Action as inspiration.

\section{\label{sec::level2} A Sample Lagrangian}
Perhaps one of the most famous analyses of human behavior comes from the great philosopher John Stuart Mill who posits a world based on Utility.
Here, we assert that any decision maker, at any given point in time, wants to maximize their utility, which we shall identify as minimizing the cost they will incur as a result of their decision.
In other terms:

$$\text{Utility} = \text{Benefit} - \text{Cost}$$

So the quantity we wish to minimize over time is:
$$ S_U = \text{Min} \int_{t_i}^{t_{i+1}} -\text{Utility}\text{ } dt= \text{Min} \int_{t_i}^{t_{i+1}} (\text{Cost} - \text{Benefit})\text{ }dt $$

The reader familiar with the Lagrangian formulation of Classical Mechanics ought to immediately recognize this as a statement of the Priciple of Least Action.
$S_U$ is analogous to:
$$S = \text{Min} \int_{t_i}^{t_{i+1}} \mathcal{L}(x,t) dt$$

So our Lagrangian ends up being $-\text{Utility}$.

\section{\label{sec::level2} Constructing the Theory using Feynman as a Guide}
The question we are now asking is identical to the one posed by Feynman.
Whereas he was concerned with the paths of particles through space-time, we are concerned with the paths of decision makers through decision space.
\par
We begin, as Feynman does, by defining the probability that the decision maker's path through decision space is contained within a region $R$ of decision space to be $|\phi(R)|$.
We call $\phi(R)$ the probability amplitude. \cite{feynman}
Those that are familiar with quantum mechanics would be advised to carefully note that $\phi(R)$ is the probability amplitude of an entire path, not a single position.
Feynman's great observation of every possible path the particle could take in the region $R$ was that:
\begin{displayquote}
	The paths contriubte equally in magnitude; but the phase of their contribution is the classical action (in units of $\hbar$); i.e., the time integral of the Lagrangian taken along the path. \cite{feynman}
\end{displayquote}
And so, the probability amplitude for our decision maker is:
$$\phi(R) = \lim_{\epsilon \to 0} \int_R \exp\left[\frac{i}{\hbar} \sum_i S_U (x_{i+1},x_i)\right]\ldots \frac{dx_{i+1}}{A} \frac{dx_i}{A} \ldots$$

where $\epsilon$ is a small step in time and for our purposes, the units of $\hbar$ are arbitrary.
\par
To complete the analogy, our next task is to reconcile this definition with our more usual definitions of wavefunctions in quantum mechanics.  
We begin by observing that no matter how the region $R$ is constructed, it must have at least one basis vector that aligns with time.  
Continuing to follow Feynman's work, we separate $R$ into two regions: $R'$, which occurs entirely in the past, and $R''$ which occurs entirely in the future.  
With a little fancy algebra, we conclude, exactly as Feynman did that ``$|\phi(R',R'')|^2$ is the (relative) probability that if the system were in region $R'$ it will be found later in $R''$''.  
Now this is starting to look more familiar.  But the usual form of quantum mechanics has a state that predicts future behaviour solely based on the past and the present.  \cite{feynman}
We take on Feynman's definition of\cite{feynman}:
$$ \phi(R',R'') = \int \chi^*(x,t)\psi(x,t) dx$$
where,
$$\psi(x_k,t) = \lim_{\epsilon \to 0} \int_{R'} \exp\left[\frac{i}{\hbar} \sum_{i=-\infty}^{k-1} S_U (x_{i+1},x_i)\right]\frac{dx_{k-1}}{A} \frac{dx_{k-2}}{A} \ldots$$
$$\chi(x_k,t)^* =   \lim_{\epsilon \to 0} \int_{R''} \exp\left[\frac{i}{\hbar} \sum_{i=k}^{\infty} S_U (x_{i+1},x_i)\right] \frac{1}{A}\frac{dx_{k+1}}{A} \frac{dx_{k+2}}{A} \ldots$$

In Feynman's own words:
``Thus, we can say: the probability that a system in state $\psi$ will be found by an experiment whose characteristic state is $\chi$ (or, more loosely, the chance that a system in state $\psi$ will appear to be in $\chi$) is:
$$ \left|\int \chi^*(x,t) \psi(x,t) dx\right|^2 \text{.'' \cite{feynman}}$$

Observe here that $\psi$ is the ``state'' the decision maker is in and the choices he or she makes are positions.  
Most importantly, they are observables.  
Now that we have defined $\psi$ and its associated probability, the traditional interpretations of quantum mechanics can take over.

\section{\label{sec::level1} Example Problem}
For clarity, we will define this problem as a financial one, where costs and benefits are more easily identifiable, but the method is easily extensible to even the most qualitative or heuristic analyses.
\par
Let's take the common problem of considering the purchase of an expensive sports car.  
For simplicity's sake, let us also assume that you must buy \underline{a} car, though it need not be the expensive sports car.

\subsection{\label{sec::level2} Decision Factors}
There are several factors to be considered in this problem.
First and most obvious is the upfront cost.
Assume the sports car costs $\$100,000$, and a reasonable sedan costs $\$50,000$
In both cases the need for a car has been fulfilled.
It seems reasonable that the benefit of having a car at all would be equal to the cost of the sedan.
For the sports car, the insurance payments are likely to be higher than for a more standard sedan, and therefore add additional costs.
Finally, there are more subjective considerations to the particular decision maker.
For instance, the percieved social or personal benefits that may come from owning the sports car over the sedan.
The more expensive car could have significantly more value to a collector than the average person.
For this example assume there are moderate social benefits to having the car, but there are no negative consequences to having the sedan.

\subsection{\label{sec::level2} Construct Lagrangian}
Now, construct the Lagrangians:
$$\mathcal{L}(\text{Car}) = \text{Sticker} + \text{Insurance} - \text{Social Benefit} - \text{Car Benefit}$$
$$\mathcal{L}(\text{Sports Car}) = 100,000 + 500 - 200 -50,000 = 50,300$$
$$\mathcal{L}(\text{Sedan}) = 50,000 + 200 - 0 - 50,000 = 200$$

Note that there are no units or arbitrary units.  
This is intentional and actually useful.  
Dollars do not represent physical attributes and the value of money is a subjective attribute.
This allows us to account for the subjective value determinations made by the decision maker.
In other words, the Lagrangians are specific to a particular decision maker.

\subsection{\label{sec::level2} Calculate States}
Using our definitions of $\psi$ and $\chi$ above we have:
$$ \chi(\text{Sports Car} ) = \exp {\left( i \mathcal{L}(\text{Sports Car})\right)} = \exp{\left(i \times 50,300\right)}  $$
$$ \chi(\text{Sedan})        =  \exp{\left( i \mathcal{L}(\text{Sedan})\right)}  = \exp{\left(i \times 200\right)} $$
$$ \psi(\text{Initial State}) = \chi(\text{SportsCar}) * \chi(\text{Sedan})$$

Now from this point, our physical intuition tells us that one will most likely buy the sedan.
In the final analysis, according to the Copenhagen Interpretation, this is exactly what happens.
However, in more complex circumstances it may not be so clear.
Furthermore, these numbers are heavily dependant on what the percieved social and personal benefits are.
These are very subjective numbers that depend heavily on the decision maker.
These calculations give

$$ P(\chi(\text{Sports Car} )) =   0.002090$$
$$ P(\chi(\text{Sedan}) )       =   0.997910$$
$$ P(\psi(\text{Initial State})) = 1$$

If we repeat these calculations with a higher personal value, $\$10,000$, placed on owning the sports car the numbers change dramatically.

$$ P(\chi(\text{Sports Car} )) =   0.561141$$
$$ P(\chi(\text{Sedan}) )       =   0.438859$$
$$ P(\psi(\text{Initial State})) = 1$$

This shows how the model is responsive to the subjective whims of the particular decision maker being modelled.

\section{\label{sec::1} Conclusion}
We can see for the example of the sports car that the model behaves as expected.
The more value the decision maker places on something, the more likely they are to choose it.
The value here is subjective, specific to the person making the decision.
Most of all however, it is a much more direct and precise analogy to the structure of quantum mechanics, allowing for more sensible interpretation.
\par
Future efforts could consist of generating full visualizations of decision spaces in the complex plane, adding uncertainties to Lagrangians, or even having multiple time steps representing multiple paths through multiple choices.
Another particularly interesting idea would be to use techniques usually associated with Machine Learning to derive the Lagrangians for key decision makers or decision making bodies.
\par
This theory still needs to be extensively experimentally tested and furthermore, there remain significant questions as to the proper normalization of the Lagrangians and adopted value metrics.
Nonetheless, with further development, it is believed this model could be improved and expanded upon to devlop a full Theory of Decision Making.





\newpage


\begin{thebibliography}{9}

\bibitem{feynman} 
Richard Feynman
\textit{Space-Time Approach to Non-Relativistic Quantum Mechanics}. 
Reviews of Modern Physics, 20(2):367-387, 1948.
 
 \bibitem{bagarello}
 Bagarello, Fabio and Basieva, Irina and Khrennikov, Andrei. 
\textit{Quantum field inspired model of decision making: Asymptotic stabilization of belief state via interaction with surrounding mental environment.}
Journal of Mathematical Psychology. vol 82:159-168, 2018 


\bibitem{ashtianai}
Mehrdad Ashtiani and Abdollahi Azgomi
\textit{A Survey of Quantum Like Approaches to decision making and cognition}
Mathematical Social Sciences, vol 25, pgs 49-80, 2015

\end{thebibliography}
\end{document}